\documentclass[aps,pra,twocolumn,groupedaddress]{revtex4}
\usepackage{latexsym}
\usepackage{amssymb}
\usepackage{amsmath}
\usepackage{amsthm}
\usepackage{amsfonts}
\usepackage{amssymb} 
\usepackage{graphicx}
\usepackage{booktabs}
\usepackage{subfigure}
\usepackage{color}

\newcommand{\be}{\begin{equation}}
\newcommand{\ee}{\end{equation}}
\newcommand{\bes}{\begin{equation*}}
\newcommand{\ees}{\end{equation*}}
\newcommand{\eqn}{\begin{eqnarray}}
\newcommand{\feqn}{\end{eqnarray}}
 
\DeclareMathOperator{\ree}{Re}
\DeclareMathOperator{\csch}{csch}

\begin{document}

\title{Emission of correlated photon pairs from superluminal perturbations in dispersive media.}

\author{F.~Dalla~Piazza,$^{1}$ F.~Belgiorno,$^{3}$ S.~L.~Cacciatori,$^{1,2}$   and D. Faccio$^{1,4}$ }
\address{$^{1}$ Dipartimento di Scienza e Alta Tecnologia, Universit\`a dell’Insubria, Via Valleggio 11, IT-22100 Como, Italy,\\
$^{2}$ I.N.F.N., Sezione di Milano, Via Celoria 16, 20133, Milano, Italy\\
$^{3}$ Dipartimento di Matematica, Politecnico di Milano, Piazza Leonardo 32, 20133 Milano, Italy\\
$^{4}$ School of Engineering and Physical Sciences, SUPA, Heriot-Watt University,
Edinburgh EH14 4AS, UK
}

\date{\today}

\begin{abstract}
We develop a perturbative theory that describes a superluminal refractive perturbation propagating in a dispersive medium  and the subsequent excitation of the quantum vacuum zero-point fluctuations. We find a process similar to the anomalous Doppler effect: photons are emitted in correlated pairs and mainly within a Cerenkov-like cone, one on the forward and the other in backward directions. The number of photon pairs emitted from the perturbation increases strongly with the degree of superluminality and under realizable experimental conditions, it can reach up to $\sim10^{-2}$ photons per pulse. Moreover, it is in principle possible to engineer the host medium so as to modify the effective group refractive index. In the presence of ``fast light'' media, e.g. a with group index smaller than unity, a further $\sim10\times$ enhancement may be achieved and the photon emission spectrum is characterized by two sharp peaks that, in future experiments would clearly identify the correlated emission of photon pairs.
\end{abstract}

\maketitle

\section{Introduction}
Since the analysis of Schwinger \cite{schw} a great amount of research has been devoted to the study of particles production from the vacuum. In the seminal paper \cite{hawk}, Hawking predicted that the presence of an horizon around a black hole induces the particle production. Actually, the existence of the Hawking radiation does not require a gravitational collapse, but rather the key elements are a quantum field and an event horizon associated with a curved space time metric \cite{unruh,visser,ulf,faccio,rubino}.
Another mechanism providing particle production is the dynamical Casimir effect \cite{fulldav}. Here, a two dimensional quantum theory of a massless scalar field is considered that is influenced by the motion of a perfectly reflecting boundary (mirror). The vacuum expectation value of the energy-momentum tensor for an arbitrary mirror trajectory shows a non vanishing radiation flux.

\begin{figure}[b]
\centering
\includegraphics[width=7cm]{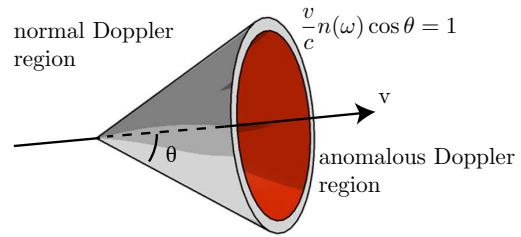}
\caption{Schematical representation of normal and anomalous Doppler emission from a ``source'' moving with velocity $v$.}
\label{fig:cerenkov}
\end{figure}

Here we consider a further effect by which particle production is induced by the superluminal motion of a perturbation of the refractive index in a dielectric medium with dispersion. Differently from the Hawking effect, the production of particles is not induced by the presence of an horizon but by the fast motion of the perturbation.  Therefore, this effect can be better recognized in the context of the  anomalous Doppler effect \cite{adoppler1,adoppler2,adoppler3,adoppler4}. Following the description given by Ginzburg, light emitted by a generic moving source will be Doppler shifted according to the formula $\omega(\theta)=\omega'/\gamma |1-v/cn(\omega)\cos\theta|$ \cite{adoppler4}, where $\omega'$ is the comoving reference frame value of the emitted frequency, $\theta$ is the observation angle and $\gamma=1/\sqrt{1-v^2/c^2}$. For $v/cn(\omega)\cos\theta<1$ we have the normal Doppler effect but for $v/cn(\omega)\cos\theta>1$, i.e. for a superluminal source  we have the so-called anomalous Doppler effect (see Fig.~\ref{fig:cerenkov}). Emission of a photon on behalf of the moving source must in general correspond to a change in the internal state, e.g. a transition in energy from one level to another. In the normal Doppler regime, a positive frequency photon will thus be emitted with a spontaneous transition from an upper to a lower energy state. In much the same way, in the anomalous Doppler regime a negative frequency photon (positive frequency in the laboratory reference frame) will be spontaneously emitted in combination with a transition from a lower to a higher state. The energy required for this process is provided by the superluminal translational movement. The superluminal source will therefore continuously emit pairs of photons, one in the anomalous and the other in the normal Doppler regions \cite{adoppler4}. Anomalous Doppler emission has never actually been observed before due to the obvious difficulty in actually realising an experimental system in which to observe these effects.\\
In this work we extend the same concept proposed to generate event horizons in optical media: an intense laser pulse focused into the medium will induce a local increase in the refractive index and this create a perturbation in the dielectric medium (DP) that travels locked with the laser pulse itself \cite{ulf,rubino}. Laser pulses will typically travel at a well-defined group velocity so that this kind of DP is ideal for creating the analogue of an event horizon. However, by properly choosing how the laser pulse overlaps with the dielectric medium it is possible to generate a DP that travels with any arbitrary velocity. For example, recent experimental results clearly show that it is possible to generate isolated peaks within the laser pulse that travel faster than the vacuum speed of light, $c$ \cite{milchberg,bonaretti}. Within the context of these findings we therefore consider the problem of how  a superluminal DP interacts with the zero-point fluctuations of the surrounding quantum vacuum. This same problem was first tackled in Ref.~\cite{superl} where it was shown that, in a dispersion-less medium a superluminal DP will spontaneously emit photons that are excited from the vacuum state. The photons appear as correlated pairs and their  wavelength decreases from the far-infrared to the visible region as the DP velocity is increased from $\sim c$ to many times $c$.
Here we extend these results to a dispersive medium the results and we consider in more detail the nature of the correlated photons and possible experimental implementation. We also consider how the correlated photon pair emission may be enhanced by properly engineering the vacuum states, i.e. by using media in which the effective group velocity at the emitted frequencies is also superluminal, so-called ``fast light media'' \cite{milonni_book}.

\section{Nondispersive case}
We start by briefly summarising the main results and equations for the non-dispersive case treated in Ref.~\cite{superl}.\\
 An intense light pulse induces the perturbation of the refractive index through the optical nonlinear Kerr effect. This perturbation travels across the medium at the same speed of the light pulse. The medium was supposed to be nondispersive. The authors find that there is no pair production unless the pulse velocity $v$ satisfies the condition $v\geq \frac{c}{n_0}$, where $c$ is the velocity of light in vacuum end $n_0$ is the uniform and constant background refraction index.
The model analyzed is based on the perturbative approach introduced by Sch\"utzhold \emph{et al.} in \cite{schutz} and the interaction representation for the electromagnetic field in the presence of a dielectric constant $\epsilon(\vec{x},t)$ depending on space and time is considered. Moreover, a uniform background and a constant value $\epsilon_b=n_0^2$ for the dielectric constant are assumed. Thus, the disturbance is $\epsilon(\vec{x},t)-\epsilon_b$. Following \cite{schutz}, the interaction Hamiltonian is ${\mathcal H}_I = \xi \Pi^2$ with $\xi := \frac 12 \left(\frac{1}{\epsilon (\vec{x},t)}-\frac{1}{\epsilon_b}\right)$ and the number of particles created, labeled by the momentum $\vec{k}$ 
and the polarization $\mu$, is:
\be
N_{\vec{k}_1 \mu_1} = \sum_{\vec{k}_2} \frac{\omega_1 \omega_2}{V^2} |\tilde{\xi}
(\underline{k}_1+\underline{k}_2)|^2
[1-(\vec{e}_{\vec{k}_2} \cdot \vec{e}_{\vec{k}_1 \mu_1})^2 ],
\ee
where $\underline{k}=(\vec{k},\omega_{\vec{k}})=(\vec{k},{c k}/{n_0})$, with $k=|\vec{k}|$, 
$\vec{e}_{\vec{k}}=\vec{k}/k$, the subscripts $1$ and $2$ label the particles of the emitted couple, and the particular shape of the perturbation is contained in the function $\xi$. 
Considering a Gaussian behavior for the refraction index $n^2=\epsilon(x,t,y,z)$:
\be
n^2 = n_0^2 + 2n_0 \eta e^{-[(x-vt)^2 +y^2 +z^2]/2\sigma^2}
\ee
the expression for the number of created particles becomes\footnote{Note that in \cite{superl} there was a misprint in the numerical constant beyond the equation \eqref{Nkmu} of a factor $2^3$.}:
\begin{widetext}
\begin{align} 
\label{Nkmu}
N_{\vec k_1, \mu_1}= 2^2 \sigma^6 \frac {\pi^2}{v^2} \frac {\eta^2}{n_0^6}  \int d\vec{k}' 
\frac{\omega_1 \omega_2}{V}
e^{-{\sigma^2} |\vec k_1 +\vec k_2|^2} \left[\delta ((\vec{k}_1+\vec{k}_2)_x - \frac{c}{v n_0} (k_1+k_2) )\right]^2
[1-(\vec{e}_{\vec{k}_2} \cdot \vec{e}_{\vec{k}_1 \mu_1})^2 ].
\end{align}
\end{widetext}
The delta function gives a constraint on the possible values of the momenta for the emitted particles.

\section{Dispersive case}
\subsection{Perturbation}
When considering actual experiments, the previous model should only be considered as an approximation of a  real medium and one has to take into account the presence of the dispersion. In this section we generalize the model of \cite{superl} for dispersive media.
To introduce dispersion in the model we have to refine the theoretic approach leading to the expression for the number of emitted particles. 
Let us write the dispersive electromagnetic action as
\begin{eqnarray}
S=\frac 12 \int \vec E \cdot { \varepsilon \vec E} dx^4,
\end{eqnarray}
where $\varepsilon \vec E$ is
\begin{eqnarray}
\varepsilon \vec E (\vec r, t)=\varepsilon(\vec r, i\partial_t) \vec E(\vec r, t).
\end{eqnarray} 
We will not specify the operator $\varepsilon$ now, but we will assume for it to be a symmetric operator.
The unperturbed conjugate momentum is then
\begin{eqnarray}
\vec \Pi=\varepsilon \vec E.
\end{eqnarray}
If we set
\begin{align}
\vec E_\omega(\vec r)&=\int_{\mathbb R} e^{-i\omega t} \vec E(\vec r, t) dt, \\
\vec \Pi_\omega(\vec r)&=\int_{\mathbb R} e^{-i\omega t} \vec \Pi(\vec r, t) dt,
\end{align}
we can write
\begin{eqnarray}
\vec E= \frac 1\varepsilon \vec \Pi=\frac 1{2\pi} \int_{\mathbb R} e^{i\omega t} \frac 1{\varepsilon (\vec r, \omega)} \vec \Pi_\omega
(\vec r) d\omega.
\end{eqnarray}
We introduce a perturbation
$\varepsilon(\vec r, i\partial_t)\mapsto\varepsilon(\vec r, i\partial_t)+\delta\varepsilon(\vec r-\vec v t, i\partial_t).$
To proceed, we note that now the momentum is
\begin{eqnarray}
\vec \Pi=\varepsilon \vec E+\delta\varepsilon \vec E.
\end{eqnarray}
To describe the Hamiltonian we have to invert this operator in order to express the electric field in terms
of the momentum. This can be done perturbatively. Let us introduce an order parameter $\eta$ so that
\begin{eqnarray}
\delta\varepsilon (\vec r -\vec v t, i\partial_t)=\eta \delta\epsilon (\vec r -\vec v t, i\partial_t).
\end{eqnarray}
Next we write
\begin{eqnarray}
\vec E= \vec E_0 +\eta \vec E_1 +\ldots
\end{eqnarray}
where the dots are higher orders in $\eta$. Then, we have
\begin{eqnarray}
\vec \Pi= \varepsilon \vec E_0 +\eta (\delta\epsilon \vec E_0 +\varepsilon \vec E_1),
\end{eqnarray}
which gives
\begin{eqnarray}
&& \vec \Pi=\varepsilon \vec E_0,\\
&& \varepsilon \vec E_1=-\delta\epsilon \vec E_0.
\end{eqnarray}
We may therefore quantize the theory perturbatively, by using the unperturbed momentum operator $\vec  \Pi$
and the perturbation
\begin{eqnarray}
\delta \vec E=\eta \vec E_1=-\frac 1\varepsilon (\delta\varepsilon \vec E_0)=
-\frac 1\varepsilon (\delta\varepsilon \frac 1\varepsilon \vec \Pi).
\end{eqnarray}
The action is then
\begin{eqnarray}
&&S=\frac 12\int \vec \Pi \cdot \vec E dx^4
\cr
&&\phantom{S}= \frac 12 \int \vec \Pi \cdot \frac 1\varepsilon \vec \Pi dx^4
-\frac 12 \int \vec \Pi \cdot \frac 1\varepsilon (\delta\varepsilon \frac 1\varepsilon \vec \Pi)dx^4+\ldots
\cr
&&\phantom{S} =\frac 12 \int \vec \Pi \cdot \frac 1\varepsilon \vec \Pi dx^4
-\frac 12 \int \frac 1\varepsilon \vec \Pi \cdot \delta\varepsilon \frac 1\varepsilon \vec \Pi dx^4+\ldots
\end{eqnarray}
where the dots are higher order terms and we have used the symmetry of $\varepsilon^{-1}$ (and dropped total time derivatives).\\
Thus, the Hamiltonian density is
\begin{eqnarray}
{\mathcal H}={\mathcal H}_0 +{\mathcal H}_I,
\end{eqnarray}
where ${\mathcal H}_0$ is the unperturbed hamiltonian (with dispersion) and 
\begin{eqnarray} \label{intham}
{\mathcal H}_I:=-\frac 12 \frac 1\varepsilon\vec \Pi \cdot \delta\varepsilon \frac 1\varepsilon \vec \Pi
\end{eqnarray}
is the first order perturbation, which reduces to the hamiltonian used in Refs.~\cite{schutz,superl} in absence of dispersion.\\
The perturbative computation is then carried forward starting from this expression.

\section{The model}
Consider a perturbation of the refractive index moving in the $x$ direction with velocity $v$, i.e. $n(\omega)(\omega)+\delta n(x-vt)$ and assume that the $\omega$ dependence of the term $\delta n$ is negligible. Thus, at the first order in $\delta n$ the dielectric constant is $\epsilon(\omega)^2+2n(\omega)\delta n(x-vt)$. The perturbation term \eqref{intham} may be taken as an operator that we can write formally as 
\be \label{oper}
\mathcal{H}_I=-\frac 12 \vec E\delta\epsilon \vec E=-\vec E \left(n(i\partial_t)\delta n(x-vt)\right)^S \vec E,
\ee
where the apex $S$ implies the Weyl prescription for the maximally symmetrised operator \cite{weyl}.
The perturbative approach used here may be justified by evaluating the main parameters under investigation. We expect to observe a similar behaviour to the non-dispersive case for which there is a peak of emission for $\lambda_{\rm max}=3\sigma$ \cite{superl}, where $\sigma$ is the radius of the gaussian-shaped perturbation.   
It is reasonable to expect that the perturbative approach works if the measured wavelength $\lambda$ satisfies $\sigma/{\lambda}>>\eta$.  For example, if $\eta$ is of the order of $0.01-0.001$ the perturbative approach is justified for emitted wavelengths well below 100 nm if $\sigma\sim1$ $\mu$m.  
In this situation the operator appearing in equation \eqref{oper} can be approximated by the more tractable one $-\frac 12 \vec E\delta\epsilon \vec E=-\vec E \left(n(i\partial_t)\delta n(x-vt)\right)^S \vec E\simeq -\vec E \delta n(x-vt)n(i\partial_t) \vec E$, in which the operator $n(i\partial_t)$ acts just on the electric field $E$ and not on the variation of the refractive index $\delta n(x-vt)$.
This perturbation has the same form as the interaction Hamiltonian $\mathcal{H}_I$ considered in \cite{schutz}, cf. equation (59). In the present case $\delta n(x-vt)$ plays the role of $\xi(\vec{r},t)$ and the electric field of $\hat{\vec{\Pi}}$. These considerations allows us to apply the perturbative scheme of \cite{schutz}. The expression of the electric field $E$ is (see \cite{milonni}, equation (32),  coherently adapted to the notation of \cite{schutz}):
\begin{align} \label{expansE}
\vec E(\vec  r,t)&=i\sum_{\vec k\mu}\left(\frac{\omega}{2V}\right)^{1/2}\frac{1}{n(\omega)n_g(\omega)}\left[a_{\vec  k\mu}(t) e^{i\vec  k\cdot\vec  r} \right.\nonumber \\
&\left.-a^\dag_{\vec k\mu}(t) e^{-i\vec k\cdot \vec r}\right]\vec e_{\vec k\mu},
\end{align}
where $n_g(\omega)=\frac{d(n(\omega)\omega)}{d\omega_k}_g(\omega)+\omega\frac{dn(\omega)}{d\omega}=\frac{c}{v_g(\omega)}$ and $\underline{k}=(\vec k,\omega)=(\vec k, ck/(n(\omega))$, with $k=|\vec k|$ and $\vec e_{\vec k}=\vec k/k$.
The action of the operator $n(i\partial_t)$ on the electric field $\vec{E}$ is:
\begin{align}
n(i\partial_t)\vec{E}&(i\partial_t)\int e^{-i\omega_k t} E_{\omega}dk \cr
&=\int n(\omega)E_{\omega}dk,
\end{align}
where $E_{\omega}$ are the Fourier modes in the expansion \eqref{expansE} of the electric field.
Thus, the number of particles of momentum $\vec{k}_1$ and polarization $\mu_1$ per unit of volume generated by the disturbance $\mathcal{H}_I$ turns out to be:
\begin{align}
<&\hat{N}_{{\bf k}_1\mu_1}>=<\psi(t\rightarrow\infty)|\hat{N}_{{\bf k}_1\mu_1}|\psi(t\rightarrow\infty)>\cr
&=\int d^4x_1\int d^4 x_2 <0|\mathcal{H}_I(x_1)\hat{N}_{{\bf k}_1\mu_1}\mathcal{H}_I(x_2)|0>,
\end{align}
where $\hat{N}_{{\bf k}_1\mu_1}$ is the number operator.
Expanding the product of the fields and the number operator in the bracket with the vacuum state and defining $f(k):=\left(\frac{\omega}{2V}\right)^{1/2}\frac{1}{n(\omega)n_g(\omega)}$ we obtain the expression:
\begin{widetext}
\begin{align} \label{prima}
N_{{\bf k}\mu}=\int d^4x_1\int d^4x_2 <0|
\sum_{\begin{array}{c}
\scriptstyle \vec k_1\mu_1,\,\vec k_2\mu_2, \\
\scriptstyle \vec k_3\mu_3,\,\vec k_4\mu_4
\end{array}}
&f(k_1)f(k_2)f(k_3)f(k_4)
\hat{a}_{\vec{k}_1\mu_1}\vec{e}_{\vec{k}_1\mu_1}
\hat{a}_{\vec{k}_2\mu_2}\vec{e}_{\vec{k}_2\mu_2}
\hat{a}^\dag_{\vec{k}\mu}\hat{a}_{\vec{k}\mu}
\hat{a}^\dag_{\vec{k}_3\mu_3}\vec{e}_{\vec{k}_3\mu_3}
\hat{a}^\dag_{\vec{k}_4\mu_4}\vec{e}_{\vec{k}_4\mu_4}\cr
&\delta n(\underline{x}_1)\delta n(\underline{x}_2)n(\omega_2)n(\omega_4)e^{-i(k_2+k_1)x}e^{i(k_3+k_4)x}|0>,
\end{align}
\end{widetext}
where the argument $\underline{x}$ of $\delta n$ is a four-vector and in our particular case it depends just on $x$ and $t$ as $x-vt$.
We indicate with a tilde the four-dimensional Fourier transform $\tilde{g}(\vec{k},\frac{ck}{n_0})=\int dtdxdydz\,g(\vec{x},t)e^{i\vec{k}\cdot\vec{x}-i\frac{ck}{n_0}t}$ and 
using the commutation rules for the operators $a_k$ and $a_k^\dag$ we obtain:
\begin{align} \label{nn}
N_{{\bf k}_1\mu_1}=f(k_1)^2
\sum_{\vec k_2\mu_2}
&\tilde{\delta}n(k_1+k_2)^2f(k_2)^2\left[n(\omega_1)+n(\omega_2)\right]^2 \\ 
&(e_{\vec k_1\mu_1}\cdot e_{\vec k_2\mu_2})^2. \nonumber
\end{align}
The space time dependence of the perturbation $dn$ is of the form
$\delta n(t,x,y,z)\equiv \delta n(x-vt,y,z)$. Hence, using the variables $u=x-vt$, $w=x+vt$, the Fourier transform of $\delta n$ is
\begin{align}\label{dn}
&\tilde{\delta}n(\underline{k}_1+\underline{k}_2)=\frac {2\pi}{v}\delta\left(k_{1x}+k_{2x}-\frac{c}{v}\left(\frac{k_1}{n(\omega_1)}+\frac{k_2}{n(\omega_2)}\right)\right) \\ \nonumber
&\int du\,dy\,dz\delta n(u,y,z)e^{i\left(k_{1x}+k_{2x}\right)u+i(k_{1y}+k_{2y})y+i(k_{1z}+k_{2z})z}.
\end{align}
\begin{figure}[b]
\centering
\includegraphics[width=7cm]{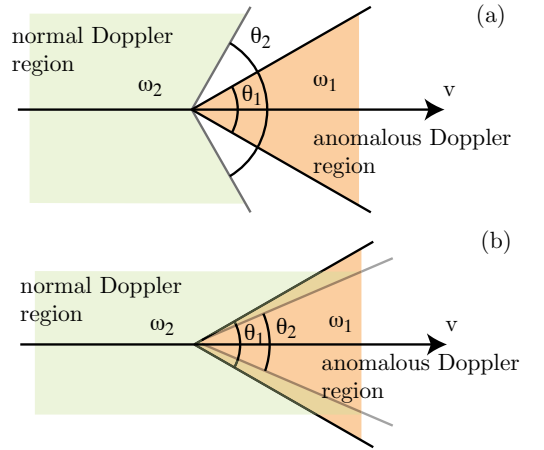}
\caption{Schematical representation of correlated photon pair emission  from a superluminal perturbation moving with velocity $v$ in the presence of dispersion: the Cerenkov cone is split in two and depending on the medium dispersion and frequency of emitted photons, regions in which no photons or both photons are emitted may appear. For the case of normal dispersion (refractive increases with increasing frequency), the figure shows the situation for $\omega_1<\omega_2$ (a) and $\omega_1>\omega_2$ (b). }
\label{fig:cones}
\end{figure}

\subsection{Superluminality emission cones}
As for the non dispersive case this last result is very meaningful from a physical point of view. The support of the delta distribution gives precise conditions for the possible states of the emitted particles. 
First we analyze the condition given by the delta function in the non dispersive case. Its argument gives the constraint:
\be
\left(k_{1x}-\frac{c}{vn_0}k_1\right) + \left(k_{2x}-\frac{c}{vn_0}k_2\right)=0.
\ee
This equation implies that  whenever $k_{1x}/k_1>c/(vn_0)$, the second photon must satisfy $k_{2x}/k_2<c/(vn_0)$, i.e. out of every pair of emitted photons, one photon is emitted inside the Cerenkov cone $\theta_0=\arccos(c/vn_0)$ (anomalous Doppler region, see Fig.~\ref{fig:cerenkov}), and the other is emitted outside the cone, in the normal Doppler region.\\
In the dispersive case we cannot always identify distinct cones of emission for the photons as before. The delta  function gives the condition:
\be
\left(k_{1x}-\frac{c}{vn(\omega_1)}k_1\right) + \left(k_{2x}-\frac{c}{vn(\omega_2)}k_2\right)=0.
\ee
Similarly to the non-dispersive case, this equation implies that  whenever  $k_{1x}/k_1>c/[vn(\omega_1)]$, the second photon must satisfy $k_{2x}/k_2<c/[vn(\omega_2)]$. 
In terms of emission angles for the two photons, we find $\theta_1<\arccos\{c/[vn(\omega_1)]\}$ and $\theta_2>\arccos\{c/[vn(\omega_2)]\}$. Therefore, if $\arccos\{c/[vn(\omega_1)]\}>\arccos\{c/[vn(\omega_2)]\}$, the two cones overlap and if $\arccos\{c/[vn(\omega_1)]\}<\arccos\{c/[vn(\omega_2)]\}$ there is a gap between them. In the first case there is a region in which both photons can be emitted (differently from the non-dispersive case), whilst in the second case there is a region in which no photon at all can be emitted. The two cases are shown in Figs.~\ref{fig:cones}(a) and (b) respectively. The non-dispersive case is obtained in the limiting situation in which $\arccos\{c/[vn(\omega_1)]\}=\arccos\{c/[vn(\omega_2)]\}$.


\subsection*{Gaussian shape of the perturbation}
We now focus attention on the problem of determining the actual number of emitted photons, starting from the assumption of a Gaussian shape for the refractive index perturbation $\delta n=\eta e^{-\frac{1}{2\sigma^2}[(x-vt)^2+y^2+z^2]}$. This form for the  perturbation is what one may expect when a local refractive index variation is induced by a laser pulse through the nonlinear Kerr effect. Indeed, laser pulses typically have Gaussian-shaped intensity, $I$, profile along both longitudinal and transverse coordinates so that $\delta n_2I$, where $n_2$ is the nonlinear Kerr index, will have the same form. A typical value for $\eta$, for example in fused silica glass, is $\sim0.001$ where $n_2\sim3\cdot 10^{-16}$ cm$^2$/W and we take $I\sim3\cdot 10^{12}$ W/cm$^2$. Inserting this $\delta n$ profile in \eqref{nn}, using $\sum_{\vec{k}} \rightarrow V\int d^3 \vec{k} \frac{1}{(2\pi)^3}$ and summing over the polarisation states $\mu_1$ and $\mu_2$ we obtain:
\begin{align}  \label{ndisp} 
&N_{{\bf k}_1}=
\frac{\sigma^6\eta^2\pi^2}{v^2V}
\frac{\omega_1}{n^2(\omega_1)n_g^2(\omega_1)}
\int d^3\vec{k}_2 \\ \nonumber
&e^{-\sigma^2|\vec k_1 +\vec k_2|^2} \frac{\omega_2}{n^2(\omega_2)n_g^2(\omega_2)} \left[n(\omega_1)+n(\omega_2)\right]^2 \\ \nonumber
&\left(\frac{k_1^2k_2^2+(\vec{k}_1\cdot \vec{k}_2)^2}{k_1^2k_2^2} \right)\left[\delta(k_{1x}+k_{2x}-\frac 1v(\omega_1+\omega_2))\right]^2. \nonumber
\end{align}
Note that in the nondispersive limit, i.e. $\frac{dn(\omega)}{d\omega}=0$, the expression for the number of emitted particles $N_{k_1}$ reduces to equation \eqref{Nkmu}.

\subsection*{Hyperbolic tangent shape of the perturbation}
We can consider another shape for the perturbation of the refractive index. In particular, let us focus on a profile given by a hyperbolic tangent function. This allows us to investigate how the particular geometry of the perturbation affects the maximum of the wavelengths and the number of emitted particles. Moreover, a $\tanh$-shape may arise if for example the dominant refractive variation is related to plasma generated through multiphoton or tunnelling ionisation by an intense laser pulse. Intense laser pulses may indeed efficiently ionise the medium within the first few optical cycles and thus create a very steep moving plasma front followed by a nearly constant plasma density that will decay (through electron recombination) over time scales of the order of $\sim100$ fs in condensed media or $\sim1$ ns in gasseous media \cite{couairon}. This situation may therefore be adequately approximated by a $\tanh$-like function. We also note that a plasma front will locally reduce the refractive index ($\delta n<0$) as opposed to the Kerr effect that will usually increase the refractive index ($\delta n>0$). Yet the result of the calculations depends not on the sign of $\delta n$ but only on its amplitude, velocity and physical dimensions.\\
The shape of the perturbation therefore is taken as:
\bes
\delta n=\eta\tanh{\frac{x-vt}{\sigma_x}}e^{-\frac{y^2}{2\sigma_y^2}}e^{-\frac{z^2}{2\sigma_z^2}}.
\ees
Again we use the variables $u=x-vt$, $w=x+vt$, and $y$ and $z$ unchanged. Repeating the same computation as above we obtain:
\begin{align} \label{nth}
N_{{\bf k}_1}&=
\frac{\sigma^2_x\sigma^2_y\sigma^2_z\eta^2\pi^2}{v^2V}\frac{\pi}{2}\frac{\omega_1}{n^2(\omega_1)n_g^2(\omega_1)}
\int d^3\vec{k}_2 \\ \nonumber
&\frac{\omega_2}{n^2(\omega_2)n_g^2(\omega_2)} \csch^2{\left[\frac{\pi\sigma_x}{2}(k_{1x}+k_{2x})\right]} \\ \nonumber
&e^{-\sigma^2_y(k_{1y}+k_{2y})^2}e^{-\sigma^2_z(k_{1z}+k_{2z})^2}  \left[n(\omega_1)+n(\omega_2)\right]^2 \\ \nonumber
&\left(\frac{k_1^2k_2^2+(\vec{k}_1\cdot \vec{k}_2)^2}{k_1^2k_2^2} \right)
\left[\delta\left(k_{1x}+k_{2x}-\frac 1v(\omega_1+\omega_2)\right)\right]^2,
\end{align}
Form these relations for the total emitted photon numbers it is immediately apparent that it is advantageous to keep a large a perturbation as possible with each of the longitudinal and transverse dimensions $\sigma_{x,y,z}$ contributing equally in the multiplicative pre-factor. As shown below, it turns out that the $\tanh$ profile emits roughly half the number of photons of a dimensionally similar Gaussian perturbation, indicating that the main role is played by the transient switch-on and off of the perturbation. We also note the dependence on the group refractive index at the emission frequency. As we shall discuss below, this allows an additional degree of freedom for enhancing or controlling the photon emission process.\\


\subsection{Numerical analysis - Gaussian shape}
The non trivial dispersion relation appearing in equation \eqref{ndisp} makes the dispersive case more intricate to analyze, even numerically, than the non dispersive one. 
We perform the numerical analysis in fused silica where, for completeness we show the full dispersion law given by the Sellmeier relation \cite{agrawal}:
\be \label{sellmeier}
n(\lambda) = \ree \left[1 + \frac{a_1\lambda^2}{\lambda^2-l_1^2} + \frac{a_2\lambda^2}{\lambda^2-l_2^2} + \frac{a_3\lambda^2}{\lambda^2-l_3^2}\right]^{1/2},
\ee
where
\begin{align*}
& a_1=0.473115591 && l_1=0.0129957170  \\ 
& a_2=0.631038719 && l_2=4.12809220\cdot 10^{-3} \\ 
& a_3=0.906404498 && l_3=98.7685322.
\end{align*}
As in the non dispersive case, the delta distribution in equation \eqref{ndisp} gives a constraint that furnishes a relation between the momenta of the produced particles. 
In principle, a component of the momentum, say $k_{2x}$, could be expressed as a function of the other ones, so that by integrating over $k_{2y}$ and $k_{2z}$ one could find explicitly the function $N_{\vec{k}_1}$ giving the distribution of the produced particles as a function of the momentum. However, for a generic dispersion law, the constraint cannot be analytically solved.
We therefore study the critical points of the integrand of equation \eqref{ndisp} that we will call $N_{k_1,k_2}$. This is a function of five variables, because the six components of the momenta are not all independent due to the constraint given by the delta. The symmetry of the problem allows to restrict the analysis to the planes $k_{1z}=0$ and $k_{2z}=0$ reducing the number of variables to three. This function describes the distribution of emitted couples as function of the momenta $k_1$ and $k_2$. Its maxima show the momenta of the couple of particles emitted with highest probability. Therefore, the effective number of emitted particles can be found integrating this distribution. 
\begin{figure}[t!]
\centering
\includegraphics{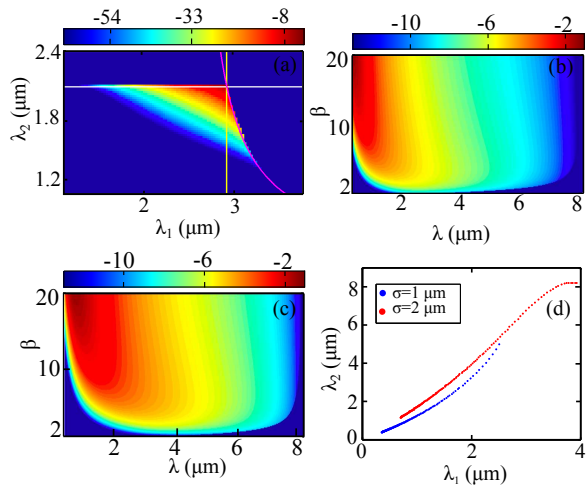}
\caption{(a) $\log_{10}N_{\lambda_1,\lambda_2}$ for $k_{1x}$ corresponding to the maximum. (b) Maxima of $\log_{10}N_{\lambda_1,\lambda_2}$ for $\theta_1=0$, $\theta_2=\pi$, and $\sigma=1 \rm{\mu\,m}$. (c) Maxima of $\log_{10}N_{\lambda_1,\lambda_2}$ for $\theta_1=0$, $\theta_2=\pi$, and $\sigma=2 \rm{\mu\,m}$. (d) Relation between the wavelengths of the emitted couples for $\beta$ increasing from 2, in the right part, to 20, in the left part of the picture.}
\label{fig:maxima}
\end{figure}
We note that the maximum of the function $N_{k_1,k_2}$ is located around $\theta_1=0$ and $\theta_2=\pi$. 
Indeed, in equation \eqref{ndisp} all factors depend on $\vec{k}_1$ and $\vec{k}_2$ by their modulus except for the terms $|\vec{k}_1+\vec{k}_2|^2$ and $(\vec{k}_1\cdot \vec{k}_2)^2$ depending on the relative orientation between the two wave vectors. For fixed $k_1$ and $k_2$ the function $N_{k_1,k_2}$ is maximum for small $|\vec{k}_1+\vec{k}_2|^2$ and large $(\vec{k}_1\cdot \vec{k}_2)^2$. 
Let be $\vec{k}_1=(k_{1x},k_{1y},0)=(k_1\cos\theta_1,k_1\sin\theta_1,0)$ and $\vec{k}_2=(k_{2x},k_{2y},0)=(k_2\cos\theta_2,k_2\sin\theta_2,0)$, then:
\begin{align*}
|\vec{k}_1+\vec{k}_2|^2 = k_1^2+k_2^2+2k_1k_2\cos(\theta_2-\theta_1)\geq &k_1^2+k_2^2 \\
-2k_1k_2 \\
(\vec{k}\cdot \vec{k}')^2 = \left[k_1k_2\cos(\theta_2-\theta_1)\right]^2 \leq k_1^2 k_2^2
\end{align*}
The two conditions are both satisfied for $\theta_2-\theta_1=\pi$, i.e. the two vectors point in opposite directions.
The condition given by the delta function in equation \eqref{ndisp} is $k_{1x}+k_{2x}-\frac cv (\frac{k_1}{n(\omega_1)}+\frac{k_2}{n(\omega_2)})=0$ that can be rewritten as $k_1\cos\theta_1-k_2\cos\theta_1-\frac cv\frac{k_1}{n(\omega_1)}-\frac cv\frac{k_2}{n(\omega_2)}=0$, where $\theta_1$ is the angle between $\vec{k}_1$ and the $x$ axis and, consequently, $\pi+\theta_2$ the angle between $k_2$ and the $x$ axis.
As observed before, the maximum of equation \eqref{ndisp} depends only on $k_1$, $k_2$ and $\vec{k}_1\cdot\vec{k}_2$,
Thus, at the maxima we have:
\begin{align*}
&\frac{d}{d\theta_1}\left[N_{k_1,k_2}(k_1,k_2,\vec{k}_1\cdot\vec{k}_2)+\lambda g(k_1,k_2,\theta_1)\right]\\
&=\frac{d}{d\theta_1}\lambda g(k_1,k_2,\theta_1)
=\lambda (k_1-k_2)\sin\theta_1=0,
\end{align*}
where $\lambda$ is a Lagrange multiplier.
Being $k_1$ and $k_2$ in general not equal and $\lambda\neq 0$, we therefore have $\theta_1=0$. We also verified numerically that indeed the maximum of $N_{k_1,k_2}$ is at $\theta_1=0$:
\begin{table}[b!]
\centering
\subtable[$\,\sigma =1$ $\mu m$\label{tab:resdisps1}]{
\begin{tabular}{cccc} 
\hline
$\beta$ & $\lambda_{1max}$ &$\lambda_{2max}$ &  $N_{\lambda_1,\lambda_2}$\\
        &  m & m & \\ 
\hline
2 & 2.51e-6 & 4.98e-6 & 6.13e-7 \\
3 & 1.93e-6 & 3.06e-6 & 6.68e-6 \\
4 & 1.54e-6 & 2.17e-6 & 3.01e-5 \\
5 & 1.26e-6 & 1.66e-6 & 9.35e-5 \\
6 & 1.08e-6 & 1.36e-6 & 2.33e-4 \\
7 & 0.94e-6 & 1.15e-6 & 5.02e-4 \\
8 & 0.83e-6 & 0.99e-6 & 8.73e-4 \\
9 & 0.75e-6 & 0.87e-6 & 1.74-3 \\
10 & 0.68e-6 & 0.78e-6 & 2.91e-3 \\
15 & 0.47e-6 & 0.52e-6 & 2.09e-2 \\
20 & 0.36e-6 & 0.39e-6 & 8.19e-2 \\
\hline
\end{tabular}
}
\subtable[$\, \sigma =2$ $\mu m$\label{tab:resdisp2}]{
\begin{tabular}{cccc} 
\hline
$\beta$ & $\lambda_{1max}$ &$\lambda_{2max}$ &  $N_{\lambda_1,\lambda_2}$\\
        & m & m & \\ 
\hline
2 & 3.93e-6 & 7.02e-6 & 4.14e-8 \\
3 & 3.50e-6 & 5.38e-6 & 2.14e-6 \\
4 & 2.95e-6 & 4.11e-6 & 1.27e-5 \\
5 & 2.49e-6 & 3.26e-6 & 4.28e-5 \\
6 & 2.13e-6 & 2.68e-6 & 1.11e-4 \\
7 & 1.87e-6 & 2.69e-6 & 2.44e-4 \\
8 & 1.65e-6 & 1.96e-6 & 4.80e-4 \\
9 & 1.48e-6 & 1.73e-6 & 8.69e-4 \\
10 & 1.35e-6 & 1.54e-6 & 1.47e-3 \\
15 & 0.92e-6 & 1.00e-6 & 1.11e-2 \\
20 & 0.70e-6 & 0.75e-6 & 4.63e-2 \\
\hline
\end{tabular}
}
\caption{\label{table} Maxima of $N_{\lambda_1,\lambda_2}$ for increasing $\beta$ and two values of $\sigma$.}
\end{table}
We visualize this in Figure \ref{fig:maxima}(a), that shows an example ($\sigma=2$ and $\beta=5$) of the distribution of photon numbers in Log scale. 
The two straight lines correspond to the condition $k_1^2=k_{1x}^2$ and $k_2^2=k_{2x}^2$ and the curve represents the relation between $\lambda_1$ and $\lambda_2$ given by the delta function in Eq.~\eqref{ndisp}. As discussed, these three curves intersect at the maximum of $N_{k_1,k_2}$ implying that indeed maximum emission occurs along the propagation direction $\theta_1=0$.\\ 
We perform the numerical analysis for perturbations with radius $\sigma=1$ $\mu$m and $\sigma=2\, \mu$m and for increasing values of $\beta$. As explained, to give an estimation of the number of emitted couples for every single DP we have to evaluate the integrand of the equation \eqref{ndisp}, i.e. $N_{k_1,k_2}:=dN/(d\Omega d\Omega 'dkdk')$. In the computation the value of $\delta(0)$ has been approximated with $L/(2\pi)$ [see Eq.~\eqref{dn}]. In the following we report all photon numbers in function of wavelength rather than wave-vector (thus indicated as $N_{\lambda_1,\lambda_2}$) and integrated over the azimuthal angles $\varphi_1$ and $\varphi_2$.

Figures \ref{fig:maxima}(b) and (c) show the calculated photon numbers $N_{\lambda_1,\lambda_2}$ in Log scale for the value of $\lambda_2$ at which emission is maximum and for $\eta=0.001$, two different values of the perturbation size, $\sigma=1$ and 2 $\mu$m respectively. Figure~\ref{fig:maxima}(d) shows both wavelengths $\lambda_1=\lambda_{1max}$ and $\lambda_2=\lambda_{2max}$ at which emission is maximum and for $2<\beta<20$. At lowest $\beta$,  $\lambda_1$ and $\lambda_2$ have the largest values and monotonically decreases and $\beta$ increases. These points are also shown for the two values of $\sigma$ considered, 1 $\mu$m (blue dots) and 2 $\mu$m (red dots). We observe that the photon emission is peaked at wavelengths $\lambda_{1max}$ that increases with the perturbation diameter. We also note that, as in the non dispersive case \cite{superl}, $\lambda_{1max}$ and $\lambda_{2max}$ decrease from the mid-infrared region in to the visible region as $\beta$, i.e. the perturbation velocity increases. However, differently from the non-dispersive case, we underline that in general $\lambda_{1max}$ and $\lambda_{2max}$ are different, i.e. the photons are correlated but at different wavelengths. However, for increasing $\beta$ this difference becomes smaller with $\lambda_{1max}\rightarrow\lambda_{2max}$ as $\beta\rightarrow\infty$.\\
Some numerical values of $N_{\lambda_1,\lambda_2}$ from these graphs are given in Table \ref{table}\ref{tab:resdisps1} and \ref{tab:resdisp2}.
An estimation of the actual number of emitted particles can be obtained by integrating the function $N_{\lambda_1,\lambda_2}$ over $\lambda$, $\theta_1$ and $\theta_2$: 
this furnishes, for $\eta=0.001$, $\sigma=1\, \rm{\mu m}$, $\beta=20$ and a 5 cm propagation distance, a number of emitted photons for every DP in an angle of 30 degree of $\sim3\cdot10^{-4}$.

\subsection{Numerical analysis - Hyperbolic tangent shape}
We perform a similar numerical analysis for a hyperbolic tangent shape of the perturbation of the refractive index by studying the integrand of equation \eqref{nth}. Adapting the previous argument to this case we search its maxima in the direction of the propagation of the DP, i.e. for $\theta=0$.
\begin{figure}[htbp]
\centering
\includegraphics[width=.48\textwidth]{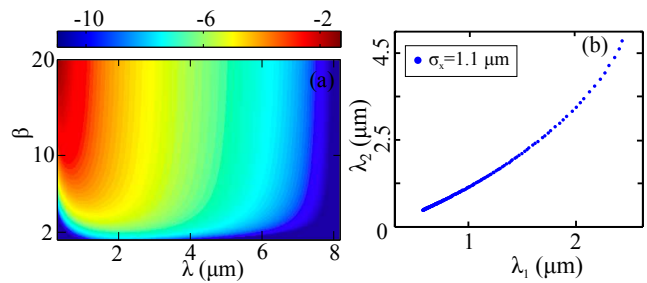}
\caption{(a) $\log_{10}N_{\lambda_1,\lambda_2}$ for $\theta_1=0$, $\theta_2=\pi$ and for a hyperbolic tangent shape of the perturbation. (b) Correlated photon wavelengths for $\beta$ increasing from 2 (top right hand point) to 20 (bottom left point).}
\label{fig:maximath}
\end{figure}
We assume a perturbation with dimensions $\sigma_x=1.1\,\rm{\mu m}$ and $\sigma_y=\sigma_z=1\, \rm{\mu m}$ and {$\eta=0.001$}.
An estimation similar to the gaussian case, for $\beta=20$ and a 5 cm propagation distance, gives a number of emitted photons for every DP in an angle of 30 degree of $\sim 1.5\cdot10^{-4}$.


\subsection{Fast light}
\begin{figure}[h]
\centering
\includegraphics{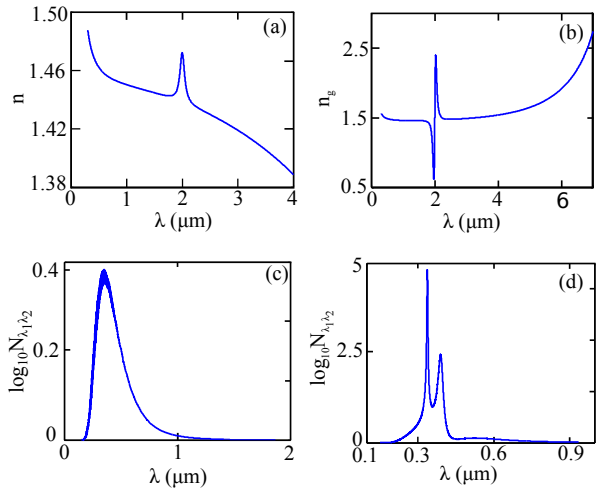}
\caption{Dispersion law (a) and group index (b) for the modified dispersion law of the fused silica. (c) $\log_{10}N_{\lambda_1,\lambda_2}$ for the standard dispersion law of the fused silica. (d) $\log_{10}N_{\lambda_1,\lambda_2}$ for the modified dispersion law of the fused silica.}
\label{fig:Nfast}
\end{figure}

We now return to a feature outlined earlier on, i.e. the dependence of the photon emissivity on the group index at the emitted frequencies [see e.g. Eq.~\eqref{ndisp}]. From our equations it is clear that by a reduction in the group velocity at $\omega_1$ and/or $\omega_2$ may greatly influence the actual number of emitted photons at both frequencies.   Group velocities exceeding the speed of light have been observed in several experiments \cite{galisteo, wang,boyd,solli,wangNature} whereby the medium is either chosen so as to have an absorption resonance close to the frequency of interest or is structurally modified, e.g. into Bragg grating structures so that the dispersion curve is strongly modified (without absorption). 
This property  can therefore be exploited to effectively engineer the (dispersion properties of the) vacuum states, increasing the number of emitted photons and, moreover, providing an additional tool for investigating the correlation properties of the produced couples.
We emulate such superluminal group velocities by introducing a simple Lorentzian correction on top a background Sellmeier relation. We therefore obtain a dispersion law with a narrow peak whose maximum slope is placed in correspondence to the point of the highest photon emission for a medium characterised only by the background dispersion.
We give an example of such a modified dispersion law for the fused silica in Fig.~\ref{fig:Nfast}(a) and the corresponding group index $n_g(\omega)(\lambda(\omega))-\lambda(\omega)\frac{dn(\lambda)}{d\lambda}$ in Fig.~\ref{fig:Nfast}(b).
We then perform the analysis for a DP of Gaussian shape with {$\eta=0.001$}, $\sigma=1\, \rm{\mu m}$ and traveling at $\beta=20$ in such a medium. In Figure \ref{fig:Nfast}(c) we show the distribution of the emitted couples in a medium with the background dispersion law and in Figure \ref{fig:Nfast}(d) the distribution produced by the  same perturbation in a medium with the Lorentzian correction to the dispersion.
\begin{figure}[t!]
\centering
\includegraphics{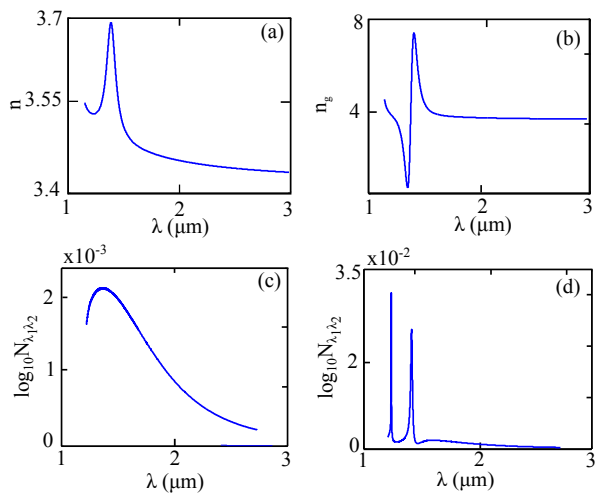}
\caption{Dispersion law (a) and group index (b) for the modified dispersion law of the silicon. (c) $\log_{10}N_{\lambda_1,\lambda_2}$ for the standard dispersion law of the silicon. (d) $\log_{10}N_{\lambda_1,\lambda_2}$ for the modified dispersion law.}
\label{fig:NfastSil}
\end{figure}
As a first comment, we observe that the number of emitted particles increases by a factor ten due to the dependence of Eq.~\eqref{ndisp} on the group indices $n_g(\omega_1)$ and $n_g(\omega_2)$. Moreover, the emission spectrum is strongly distorted but most importantly, two distinct maxima appear. These two peaks are related to enhanced emission separately at $\lambda_1$ or at $\lambda_2$. In virtue of the nature of the two emitted photons, enhancement at one wavelength will necessarily lead to enhancement at the correlated photon wavelength. Therefore, observation of these two emission peaks in an experiment could be considered as evidence of correlated photon emission from the superluminal perturbation by measuring only in the forward direction.  We performed a similar analysis for the case of silicon as this is a widely used material in optics and waveguide technology where both slow and fast light may be engineered. 
We obtain a similar increase of the emitted particles as shown in Fig.~\ref{fig:NfastSil} for a perturbation of dimension $\sigma=1\, \rm{\mu m}$ and velocity $\beta=4$.
\begin{figure}[t!]
\centering
\includegraphics[width=5cm]{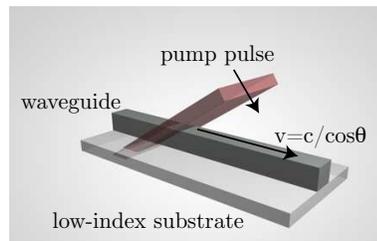}
\caption{Schematical representation of an experimental layout for creating a superluminal refractive index perturbation in a waveguide.}
\label{fig:experiment}
\end{figure}

\subsection{Conclusions}
Perturbative analysis of quantum fluctuation excitation from a travelling refractive index perturbation indicates that emission of correlated photon pairs occurs only if $v>c/n(\omega)$. This effect bears a strong resemblance to the anomalous Doppler effect and in this sense represent the first detailed analysis of a setting in which the effect may actually be observed. The number of emitted photon pairs is relatively small close to the threshold $v=c/n(\omega)$ but increases significantly if $v$ is much larger (e.g. $10-20\times$) larger than $c$. Our analysis fully accounts for material dispersion and, doing so, introduces an intriguing dependence on the group index or group velocity of the emitted photons. This dependence and careful engineering of the host medium may be used to enhance photon emission by at least an order of magnitude and also provides direct evidence of correlated emission in the form of two correlated peaks in the output spectra. \\
From our analysis it would seem that experiments are indeed feasible. Refractive index perturbations of the order of those used in this work, $\eta\sim0.001$, may be obtained by focusing laser pulses in fused silica or other nonlinear media and higher values have been observed. According to our results this would result in a photon pair emission rate of $\sim10^{-6}$ pairs/pulse at $\beta=20$ in a 5 mm long waveguide, i.e. $\sim1$ photon/second if a MHz repetition-rate laser is used. These numbers may be increased by an order of magnitude or more e.g. by using fast light media or higher DP amplitudes. The high velocities used here for the perturbation may be obtained experimentally by sending an extended (approximately ``plane wave'') laser pulse at an angle onto the host medium that could be, for example, a waveguide (see Fig.~\ref{fig:experiment}). The pulse would intersect the waveguide with an angle $\theta$ and the Kerr effect would create a DP only at the intersection point that travels with speed $c/\cos\theta$ (if we approximate the refractive index of air $\sim1$). Therefore, an incidence angle of $\sim80-85$ degrees (i.e. close to normal incidence on the waveguide) would guarantee $\beta\sim5-20$. We note that a similar approach has actually been used in previous pioneering  experiments in which a DC field was converted to  TeraHertz radiation by a superluminal laser pulse excitation of an array of biased capacitors \cite{zigler}. In our case we are exciting a superluminal DP in a dielectric medium and correlated photon pairs would be generated and collected from the two ends of the waveguide.

\subsection*{Acknowledgments}
The authors thank T. Krauss and G. Ortenzi for fruitful discussions. F. D.P. acknowledges financial support from l'Universit\'a dell'Insubria and Fondazione Cariplo, Univercomo and Banca del Monte di Lombardia.

\end{document}